\documentclass[journal=acsnano, manuscript=article, layout=twocolumn]{achemso}
\usepackage[english]{babel}

\usepackage{amsmath}
\usepackage{graphicx}
\usepackage{caption}
\usepackage{natbib}
\usepackage{siunitx}
\usepackage{booktabs}
\usepackage[colorlinks=true, allcolors=blue]{hyperref}
\usepackage{multirow} % 表格跨行必备
\usepackage{amsmath}  % 数学符号、下标必备
\usepackage{makecell}

\title{Interface-Engineered Giant Multistate Resistance Switching in Altermagnetic CrSb Multiferroic Tunnel Junctions}

  \author{Zhi Yan}
 \email{yanzhi@sxnu.edu.cn}
\affiliation{School of Materials Science and Engineering $\&$ Key Laboratory of Magnetic Molecules and Magnetic Information Materials of Ministry of Education, Shanxi Normal University, Taiyuan 030031, China}
\alsoaffiliation{Research Institute of Materials Science $\&$ Shanxi Key Laboratory of Advanced Magnetic Materials and Devices, Shanxi Normal University, Taiyuan 030031, China}
  \author{Yuwen Hua}
\affiliation{School of Materials Science and Engineering $\&$ Key Laboratory of Magnetic Molecules and Magnetic Information Materials of Ministry of Education, Shanxi Normal University, Taiyuan 030031, China}
  \author{Yueting Li}
\affiliation{School of Materials Science and Engineering $\&$ Key Laboratory of Magnetic Molecules and Magnetic Information Materials of Ministry of Education, Shanxi Normal University, Taiyuan 030031, China}
  \author{Xujin Zhang}
\affiliation{School of Materials Science and Engineering $\&$ Key Laboratory of Magnetic Molecules and Magnetic Information Materials of Ministry of Education, Shanxi Normal University, Taiyuan 030031, China}
  \author{Jianhua Xiao}
\affiliation{School of Materials Science and Engineering $\&$ Key Laboratory of Magnetic Molecules and Magnetic Information Materials of Ministry of Education, Shanxi Normal University, Taiyuan 030031, China}
  \author{Xiaohong Xu}
 \email{xuxh@sxnu.edu.cn}
\affiliation{School of Materials Science and Engineering $\&$ Key Laboratory of Magnetic Molecules and Magnetic Information Materials of Ministry of Education, Shanxi Normal University, Taiyuan 030031, China}
\alsoaffiliation{Research Institute of Materials Science $\&$ Shanxi Key Laboratory of Advanced Magnetic Materials and Devices, Shanxi Normal University, Taiyuan 030031, China}

\begin{document}

\begin{abstract}
Altermagnets enable spin-split transport without stray magnetic fields, yet converting their momentum-dependent spin splitting into a strong tunnel-junction response requires interface-selected tunneling channels. Here, using density functional theory combined with nonequilibrium Green's function calculations, we demonstrate giant multistate resistance switching in CrSb/$\alpha$-In$_2$Se$_3$ altermagnetic multiferroic tunnel junctions. The response is governed not by the bulk spin splitting of CrSb alone, but by a symmetry-selected interfacial mechanism in which Cr/Sb terminations and \textit{h}-BN or graphene insertion layers determine spin-channel matching, while ferroelectric polarization reshapes the electrostatic barrier. Symmetric and asymmetric terminations reverse the correspondence between parallel/antiparallel N\'eel-vector configurations and high-/low-resistance states, showing that the actual alignment of interfacial Cr moments selects the dominant tunneling channels. Monolayer-In$_2$Se$_3$ junctions exhibit four nonvolatile resistance states, with tunneling magnetoresistance (TMR) and tunneling electroresistance (TER) reaching 1626\% and 2206\%, respectively, and increasing to 9576\% and 4144\% upon Fermi-level shifting. Finite-bias calculations further reveal robust spin filtering and tunable spin-polarized currents. Extending the barrier to bilayer In$_2$Se$_3$ introduces interlayer polarization coupling, enabling eight resistance states with maximum TMR and TER values of $3.77\times10^{4}\%$ and $4.18\times10^{5}\%$, respectively. These results establish interface symmetry, spin-channel matching, and ferroelectric barrier reconstruction as design principles for stray-field-free multistate spintronic tunnel devices.
\end{abstract}

\vspace{1em}

\begin{tocentry}
\centering
\includegraphics[width=8.2cm]{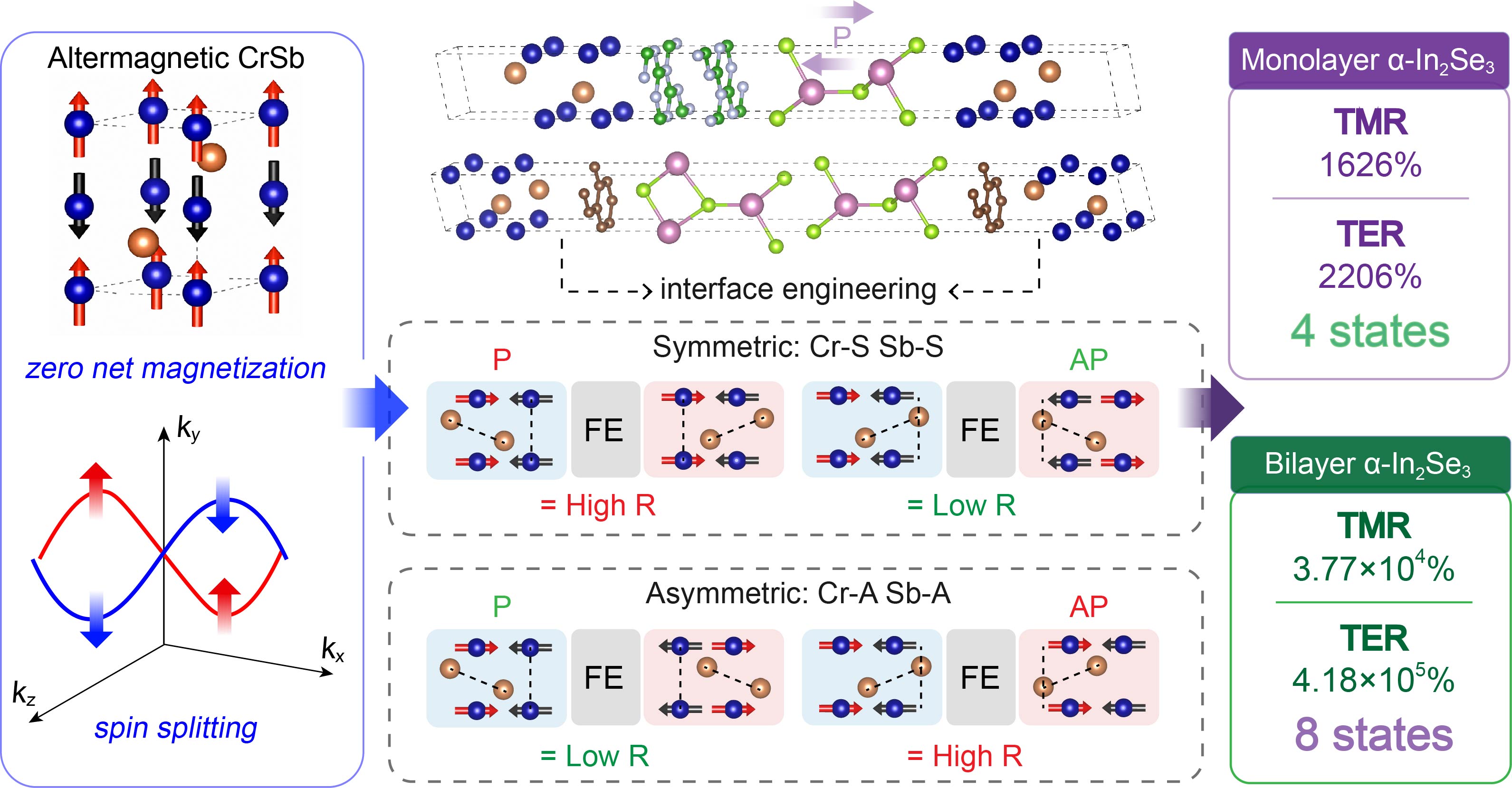}
\end{tocentry}

\maketitle

\section{Introduction}

Spintronic devices exploit the electron spin degree of freedom for information storage and processing and are widely regarded as promising candidates for low-power information technologies \cite{rotjanapittayakul2018scirep,chappert2007naturemater}. Among them, multiferroic tunnel junctions (MFTJs), consisting of two magnetic electrodes separated by a thin ferroelectric barrier, are of particular interest because they combine tunneling magnetoresistance (TMR) and tunneling electroresistance (TER) in a single device platform \cite{fusil2014annurevmater}. By independently switching the ferroelectric polarization and the magnetic configuration of the electrodes, MFTJs can in principle realize multiple nonvolatile resistance states \cite{zhuravlev2010physrevb,yan2025nanolett,yan2025physrevb2,garcia2014natcommun,cao2020acsami,wu2024nanolett,yan2026twodimensional,yan2024giant}. However, conventional MFTJs usually rely on ferromagnetic electrodes, whose stray fields and relatively limited dynamical response can restrict device stability, scalability, and operation speed. Antiferromagnets avoid these drawbacks because of their compensated magnetization and ultrafast spin dynamics \cite{baltz2018revmodphys,xiong2022fundamres,shao2024npjspintronics}, but their spin-degenerate electronic structures generally suppress spin-polarized tunneling and have limited their use in tunnel-junction devices \cite{baldrati2018physrevb,fischer2018physrevb}. Thus, an outstanding challenge is to realize strong spin-selective tunneling in a magnetic system that remains free of net magnetization.

Altermagnets provide a promising route to address this challenge. Although they possess compensated collinear magnetic order, the opposite-spin sublattices are connected by rotation, mirror, glide, or screw symmetries rather than by inversion or simple translation \cite{naka2019natcommun,smejkal2022physrevx}. This symmetry structure produces momentum-dependent spin splitting in the absence of net magnetization, with characteristic $d$-, $g$-, or $i$-wave forms \cite{fernandes2024physrevb,cheng2024physrevb,ghorashi2024physrevlett,gonzalezbetancourt2023physrevlett}. Consequently, altermagnets combine the stray-field-free character of antiferromagnets with spin-selective transport features usually associated with ferromagnets \cite{smejkal2022physrevx2,zhang2024physrevlett,fernandes2024physrevb,cheng2024physrevb,ghorashi2024physrevlett}. A variety of time-reversal-symmetry-breaking responses have been predicted or observed in altermagnets, including anomalous Hall and Nernst effects, spin-split torques, and tunneling magnetoresistance \cite{gonzalezbetancourt2023physrevlett,hariki2024physrevb,badura2025natcommun,han2025physrevappl,karube2022physrevlett,gonzalezhernandez2021physrevlett,naka2019natcommun,naka2021physrevb}. In particular, large TMR has been predicted in tunnel junctions with altermagnetic electrodes when the relative orientation of the Néel vectors is varied \cite{Zhang2025PRBCrSbMFTJ,Samanta2024PRB,Liu2024PRB}. Experimentally, spin-split altermagnetic band structures have been resolved in CrSb, MnTe, and KV$_2$Se$_2$O \cite{yang2025natcommun,lee2024physrevlett,jiang2025natphys}. Among these candidates, metallic CrSb is especially attractive for tunnel-junction applications because it combines a high Néel temperature of approximately 700 K with a large experimentally observed spin splitting of 0.93 eV \cite{ding2024physrevlett}.

These properties make CrSb a natural electrode candidate for altermagnetic MFTJs. Once a ferroelectric barrier is introduced, however, the resistance response is no longer determined solely by the Néel-vector configuration, the magnitude of the altermagnetic spin splitting, or the direction of ferroelectric polarization. At the atomic scale, the electrode termination, barrier composition, and interfacial insertion geometry can modify orbital hybridization, local electrostatic potential, and spin-resolved tunneling channels, thereby affecting both TMR and TER. This interfacial sensitivity is particularly important for altermagnetic junctions, because the spin splitting is strongly momentum dependent and the vertical tunneling current is dominated by selected regions of the two-dimensional Brillouin zone. Therefore, the key issue is not simply whether CrSb can generate a large TMR, but how the interface symmetry selects the momentum- and spin-resolved channels that convert the bulk altermagnetic spin splitting into a measurable tunnel-junction response. In this sense, interface termination, spin-channel matching, and ferroelectric barrier reshaping should be treated as coupled microscopic degrees of freedom rather than as separate structural details.

\begin{figure*}[htb!]
\centering
\includegraphics[width=17cm,height=0.7\textheight,keepaspectratio]{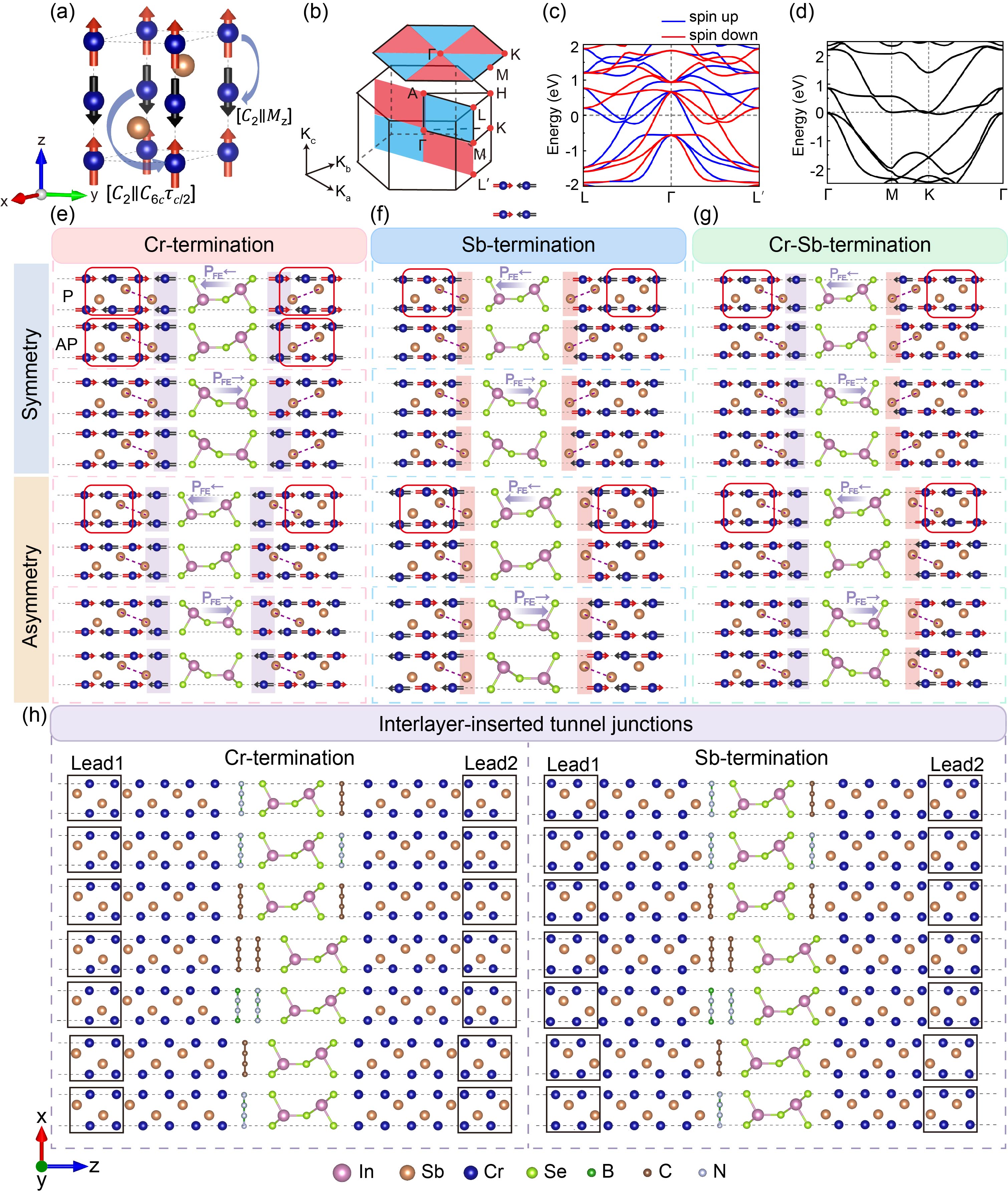}
\captionsetup{width=\textwidth}
\caption{(a) Atomic and spin configurations of CrSb, where the opposite spin sublattices are connected by rotation and mirror operations. (b) High-symmetry $k$-paths and three-dimensional Brillouin zone with colored nodal planes of CrSb. (c) and (d) Band structures of CrSb along different $k$-paths. (e--g) Schematic atomic structures of CrSb/In$_2$Se$_3$/CrSb tunnel junctions with a monolayer In$_2$Se$_3$ barrier under different interface terminations: (e) Cr-termination, (f) Sb-termination, and (g) Cr-Sb-termination. In each panel, the upper and lower parts correspond to the symmetric and asymmetric interface configurations, respectively. The purple arrows denote the ferroelectric polarization direction $P_{\rm FE}$ of the In$_2$Se$_3$ barrier, while the red and blue arrows represent the opposite spin sublattices in the CrSb electrodes. The shaded regions highlight the interfacial atomic layers.(h) Schematic atomic structures of interlayer-inserted CrSb/In$_2$Se$_3$/CrSb tunnel junctions with Cr-terminated and Sb-terminated interfaces. The black rectangles mark the periodic unit cells of the left and right electrodes, denoted as Lead1 and Lead2, respectively.}
\label{Fig1}
\end{figure*}

Here, we investigate interface-dependent multiferroic transport in CrSb/$\alpha$-In$_2$Se$_3$-based altermagnetic MFTJs using first-principles calculations combined with nonequilibrium Green's function methods. In contrast to previous CrSb-based altermagnetic tunnel-junction studies that mainly exploited Néel-vector-dependent spin splitting or ferroelectric-polarization switching, the present work focuses on how atomic-scale interface symmetry selects the active tunneling channels and couples them to ferroelectric barrier reconstruction. We consider \textit{h}-BN, graphene, and ferroelectric $\alpha$-In$_2$Se$_3$ as barrier or insertion components and construct a series of junctions with different barrier configurations and CrSb terminations. Our results reveal a symmetry-selected interface-termination--spin-channel-matching--ferroelectric-barrier-reshaping mechanism for multistate resistance switching. For monolayer-In$_2$Se$_3$ barriers, Cr-terminated interfaces generally support stronger spin-selective tunneling than Sb-terminated ones, while switching between symmetric and asymmetric terminations provides an additional route to high- and low-resistance states. In contrast, bilayer-In$_2$Se$_3$ barriers introduce interlayer ferroelectric polarization coupling and give rise to a richer dependence on polarization configuration and insertion-layer composition, enabling additional resistance states and substantially enhanced resistance modulation. These findings identify interface symmetry as a controllable microscopic variable for engineering multistate transport in altermagnetic MFTJs and suggest CrSb-based heterostructures as potential candidates for stray-field-free nonvolatile spintronic memory devices.

\section{RESULTS AND DISCUSSION}
\subsection{Structural Models of the Multiferroic Tunnel Junctions}

Hexagonal CrSb crystallizes in the NiAs-type structure with space group $P6_3/mmc$, as shown in Fig.~\ref{Fig1}(a). In this structure, each Cr atom is coordinated by six Sb atoms, and the magnetic moments of adjacent Cr layers are antiparallel along the $z$ direction, forming an A-type antiferromagnetic (A-AFM) ground state \cite{smejkal2022physrevx,park2020physrevb}. The two opposite-spin sublattices are related by the spin-group symmetry operations $[C_2\Vert M_c]$ or $\left[C_2\Vert C_{6c}\tau_{c/2}\right]$, where $C_2$ denotes spin reversal, $M_c$ is the mirror operation perpendicular to the $c$ axis, and $C_{6c}\tau_{c/2}$ represents a $60^\circ$ rotation around the $c$ axis combined with a half-unit-cell translation. These symmetry operations protect spin degeneracy on specific momentum planes while allowing momentum-dependent spin splitting away from them, giving rise to the characteristic $g$-wave altermagnetic spin splitting in CrSb \cite{ding2024physrevlett}. As illustrated in Fig.~\ref{Fig1}(b,d), the $\Gamma$-M-K-$\Gamma$ path lies within the symmetry-protected planes and therefore remains spin degenerate [Fig.~\ref{Fig1}(d)], whereas pronounced spin splitting appears along the off-plane L-$\Gamma$-L$'$ path [Fig.~\ref{Fig1}(c)]. This splitting mainly originates from exchange interactions rather than spin-orbit coupling and provides the spin-selective electronic structure required for altermagnetic tunneling \cite{smejkal2022physrevx,smejkal2022physrevx2}. These features make CrSb a suitable altermagnetic electrode for realizing spin-dependent tunneling in MFTJs.

We use monolayer $\alpha$-In$_2$Se$_3$ as the ferroelectric barrier because it is a prototypical two-dimensional ferroelectric with robust out-of-plane polarization near room temperature. The monolayer consists of five atomic layers stacked in the Se-In-Se-In-Se sequence, and its ferroelectricity mainly originates from the vertical asymmetry of the internal In-Se bonds, which gives rise to a switchable electric dipole along the $z$ direction \cite{ding2017natcommun,yan2022physrevb}. This out-of-plane switchable polarization provides the essential degree of freedom for realizing tunneling electroresistance and nonvolatile resistance switching in MFTJs. To verify that the altermagnetic character of the electrodes is preserved after forming the heterostructure, we compared the total energies of ferromagnetic and A-AFM configurations for both polarization states of CrSb/$\alpha$-In$_2$Se$_3$. The results show that the CrSb electrodes retain the A-AFM ground state in the presence of the ferroelectric barrier, as shown in Fig.~S1(a,b). Thus, the CrSb/$\alpha$-In$_2$Se$_3$ heterostructure provides the required magnetic and ferroelectric ingredients for constructing altermagnetic MFTJs.

\begin{figure*}[htb!]
\centering
\includegraphics[width=14cm,height=0.7\textheight,keepaspectratio]{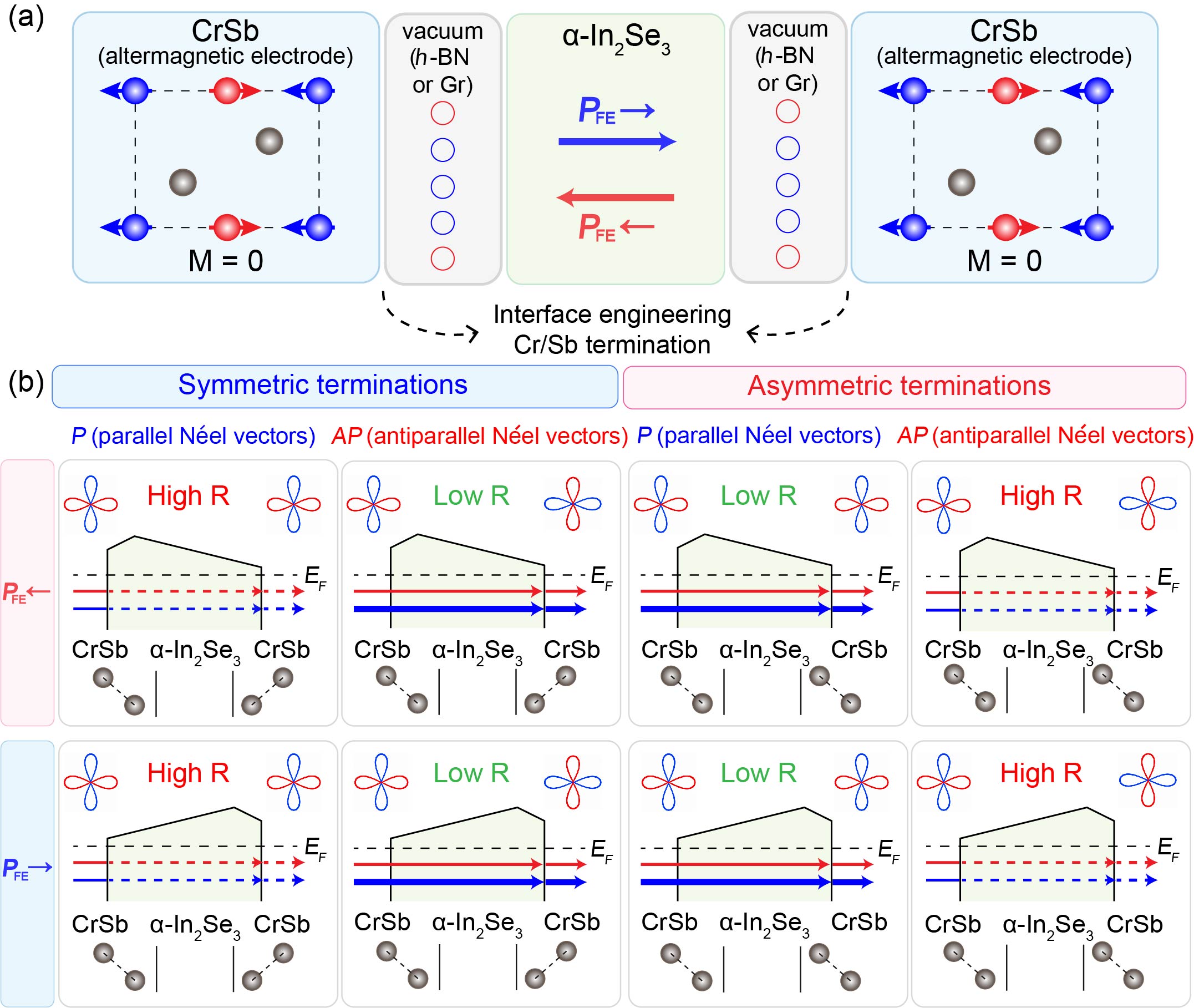}
\captionsetup{width=\textwidth}
\caption{Schematic mechanism of interface-controlled multistate resistance switching in CrSb/$\alpha$-In$_2$Se$_3$ altermagnetic multiferroic tunnel junctions. 
(a) Device structure consisting of two CrSb altermagnetic electrodes, vacuum or atomically thin \textit{h}-BN/graphene insertion layers, and an $\alpha$-In$_2$Se$_3$ ferroelectric barrier. The CrSb electrodes possess compensated altermagnetic order with zero net magnetization ($M=0$), while the ferroelectric polarization of $\alpha$-In$_2$Se$_3$ can be reversed between $P_{\rm FE}\rightarrow$ and $P_{\rm FE}\leftarrow$. The Cr/Sb termination and insertion layers provide atomic-scale interface engineering knobs for tuning spin-dependent tunneling. 
(b) Resistance hierarchy under symmetric and asymmetric interface terminations for the two ferroelectric polarization states and the parallel (P) or antiparallel (AP) N\'eel-vector configurations. The red and blue petal-like symbols represent interface-selected altermagnetic spin channels, and the tilted barrier profiles illustrate the polarization-dependent electrostatic potential. Thick solid arrows indicate favorable spin-channel matching and low-resistance states, whereas dashed arrows indicate suppressed tunneling and high-resistance states. For symmetric terminations, the P configuration gives high resistance and the AP configuration gives low resistance for both polarization directions. In contrast, asymmetric terminations reverse this correspondence. Therefore, the resistance state is governed by the actual alignment of interfacial Cr moments and the associated spin-channel matching, rather than by the nominal P/AP label alone.
}
\label{Fig2}
\end{figure*}

We first investigated CrSb/$\alpha$-In$_2$Se$_3$/CrSb multiferroic tunnel junctions with only a monolayer $\alpha$-In$_2$Se$_3$ barrier. To further investigate the effects of interface chemistry and structural symmetry, six representative interface configurations were considered, as shown in Fig.~\ref{Fig1}(e--g): Cr-S (Cr-symmetric), Cr-A (Cr-asymmetric), Sb-S (Sb-symmetric), Sb-A (Sb-asymmetric), Cr-Sb-S (Cr-Sb-symmetric), and Cr-Sb-A (Cr-Sb-asymmetric). Here, S and A denote whether the two electrode/barrier interfaces are symmetric or asymmetric with respect to the barrier center. 
In the transport calculations, the N\'eel vector of the left CrSb electrode is kept fixed, while that of the right electrode is set either parallel (P) or antiparallel (AP) to the left one. Together with the two opposite ferroelectric polarization states of $\alpha$-In$_2$Se$_3$, this setup enables both TMR and TER to be evaluated within the same junction platform. The resulting model series provides a controlled framework for clarifying how altermagnetic spin splitting, ferroelectric polarization, barrier composition, and atomic-scale interface termination jointly determine multistate tunneling transport.

\begin{table*}[htbp]
\centering
\caption{Spin-resolved transmission coefficients, TER, and tunneling magnetoresistance of different interfaces with a monolayer In$_2$Se$_3$ barrier.}
\label{tab:spin_resolved_transmission_all_interfaces}
\renewcommand{\arraystretch}{1.25}
\resizebox{\textwidth}{!}{%
\begin{tabular}{c c c c c c c c c}
\hline\hline
\multirow{2}{*}{Interface termination} & \multirow{2}{*}{\makecell{Polarization\\and Ratio}}
& \multicolumn{3}{c}{P state}
& \multicolumn{3}{c}{AP state}
& \multirow{2}{*}{TMR} \\
\cline{3-5} \cline{6-8}
& & $T_{\uparrow}$ & $T_{\downarrow}$ & $T_{\mathrm{tot}}=T_{\uparrow}+T_{\downarrow}$
& $T_{\uparrow}$ & $T_{\downarrow}$ & $T_{\mathrm{tot}}=T_{\uparrow}+T_{\downarrow}$
& \\
\hline
\multirow{3}{*}{Cr-S}
& $P_{\mathrm{FE}}\leftarrow$ & $4.04\times10^{-2}$ & $6.21\times10^{-2}$ & $1.02\times10^{-1}$ & $2.06\times10^{-2}$ & $4.83\times10^{-2}$ & $6.89\times10^{-2}$ & 48.8\% \\
& $P_{\mathrm{FE}}\rightarrow$ & $3.08\times10^{-2}$ & $6.02\times10^{-3}$ & $3.68\times10^{-2}$ & $1.39\times10^{-2}$ & $4.07\times10^{-2}$ & $5.46\times10^{-2}$ & 48.5\% \\
\cline{2-9}
& TER & \multicolumn{3}{c}{$178.5\%$} & \multicolumn{3}{c}{$26\%$} & \\
\hline
\multirow{3}{*}{Cr-A}
& $P_{\mathrm{FE}}\leftarrow$ & $2.53\times10^{-2}$ & $2.51\times10^{-2}$ & $5.05\times10^{-2}$ & $9.13\times10^{-3}$ & $3.69\times10^{-2}$ & $4.60\times10^{-2}$ & 9.6\% \\
& $P_{\mathrm{FE}}\rightarrow$ & $1.68\times10^{-2}$ & $4.84\times10^{-2}$ & $6.57\times10^{-2}$ & $6.83\times10^{-2}$ & $1.62\times10^{-2}$ & $6.45\times10^{-2}$ & 1.8\% \\
\cline{2-9}
& TER & \multicolumn{3}{c}{$30.19\%$} & \multicolumn{3}{c}{$40.1\%$} & \\
\hline
\multirow{3}{*}{Sb-S}
& $P_{\mathrm{FE}}\leftarrow$ & $1.92\times10^{-2}$ & $2.66\times10^{-2}$ & $4.58\times10^{-2}$ & $1.53\times10^{-2}$ & $2.84\times10^{-2}$ & $4.37\times10^{-2}$ & 4.8\% \\
& $P_{\mathrm{FE}}\rightarrow$ & $2.49\times10^{-2}$ & $8.30\times10^{-3}$ & $3.32\times10^{-2}$ & $1.32\times10^{-2}$ & $3.07\times10^{-2}$ & $4.39\times10^{-2}$ & 31.9\% \\
\cline{2-9}
& TER & \multicolumn{3}{c}{$37.86\%$} & \multicolumn{3}{c}{$0.3\%$} & \\
\hline
\multirow{3}{*}{Sb-A}
& $P_{\mathrm{FE}}\leftarrow$ & $1.15\times10^{-2}$ & $1.90\times10^{-2}$ & $3.05\times10^{-2}$ & $1.19\times10^{-2}$ & $2.81\times10^{-2}$ & $4.00\times10^{-2}$ & 31.32\% \\
& $P_{\mathrm{FE}}\rightarrow$ & $1.61\times10^{-2}$ & $2.34\times10^{-2}$ & $3.97\times10^{-2}$ & $2.81\times10^{-2}$ & $1.18\times10^{-2}$ & $3.99\times10^{-2}$ & 0.56\% \\
\cline{2-9}
& TER & \multicolumn{3}{c}{$30.3\%$} & \multicolumn{3}{c}{$0.2\%$} & \\
\hline
\multirow{3}{*}{Cr-Sb-S}
& $P_{\mathrm{FE}}\leftarrow$ & $4.83\times10^{-2}$ & $1.76\times10^{-3}$ & $5.01\times10^{-2}$ & $8.19\times10^{-3}$ & $3.33\times10^{-2}$ & $4.15\times10^{-2}$ & 20.7\% \\
& $P_{\mathrm{FE}}\rightarrow$ & $3.95\times10^{-2}$ & $1.36\times10^{-3}$ & $4.08\times10^{-2}$ & $1.47\times10^{-2}$ & $1.59\times10^{-2}$ & $3.06\times10^{-2}$ & 33.3\% \\
\cline{2-9}
& TER & \multicolumn{3}{c}{$22.8\%$} & \multicolumn{3}{c}{$35.6\%$} & \\
\hline
\multirow{3}{*}{Cr-Sb-A}
& $P_{\mathrm{FE}}\leftarrow$ & $3.28\times10^{-2}$ & $8.30\times10^{-3}$ & $4.11\times10^{-2}$ & $2.19\times10^{-3}$ & $4.83\times10^{-2}$ & $5.05\times10^{-2}$ & 22.9\% \\
& $P_{\mathrm{FE}}\rightarrow$ & $2.23\times10^{-2}$ & $1.51\times10^{-2}$ & $3.74\times10^{-2}$ & $1.57\times10^{-3}$ & $4.80\times10^{-2}$ & $4.96\times10^{-2}$ & 32.6\% \\
\cline{2-9}
& TER & \multicolumn{3}{c}{$9.9\%$} & \multicolumn{3}{c}{$1.8\%$} & \\
\hline\hline
\end{tabular}%
}
\label{table1}
\end{table*}

\renewcommand{\arraystretch}{1.3}
\begin{table*}[ht]
\centering
\resizebox{\textwidth}{!}{
\begin{tabular}{c c c c c c c c c}
\hline\hline
\multicolumn{9}{c}{Cr-S} \\
\hline
\multirow{2}{*}{Barrier layer} & \multirow{2}{*}{\makecell{Polarization\\and Ratio}} & \multicolumn{3}{c}{P state} & \multicolumn{3}{c}{AP state} & \multirow{2}{*}{TMR} \\
\cmidrule(lr){3-5} \cmidrule(lr){6-8}
& & $T_\uparrow$ & $T_\downarrow$ & $T_\text{tot}=T_\uparrow+T_\downarrow$ & $T_\uparrow$ & $T_\downarrow$ & $T_\text{tot}=T_\uparrow+T_\downarrow$ & \\
\hline
\multirow{3}{*}{\textit{h}-BN/In$_2$Se$_3$}
& $P_{\mathrm{FE}}\rightarrow$ & $1.32\times10^{-3}$ & $1.15\times10^{-2}$ & $1.28\times10^{-2}$ & $2.19\times10^{-3}$ & $4.05\times10^{-2}$ & $4.27\times10^{-2}$ & 234\% \\
& $P_{\mathrm{FE}}\leftarrow$ & $3.04\times10^{-3}$ & $7.42\times10^{-3}$ & $1.05\times10^{-2}$ & $1.76\times10^{-3}$ & $4.17\times10^{-3}$ & $5.93\times10^{-3}$ & 77\% \\
\cline{2-9}
& TER & \multicolumn{3}{c}{22\%} & \multicolumn{3}{c}{620\%} & \\
\hline
\multirow{3}{*}{Gr/In$_2$Se$_3$}
& $P_{\mathrm{FE}}\rightarrow$ & $2.43\times10^{-3}$ & $9.97\times10^{-3}$ & $1.24\times10^{-2}$ & $2.79\times10^{-3}$ & $1.95\times10^{-3}$ & $4.74\times10^{-3}$ & 162\% \\
& $P_{\mathrm{FE}}\leftarrow$ & $2.29\times10^{-3}$ & $1.03\times10^{-2}$ & $1.26\times10^{-2}$ & $2.46\times10^{-3}$ & $3.74\times10^{-3}$ & $6.20\times10^{-3}$ & 103\% \\
\cline{2-9}
& TER & \multicolumn{3}{c}{1.6\%} & \multicolumn{3}{c}{30.8\%} & \\
\hline
\multirow{3}{*}{\textit{h}-BN/In$_2$Se$_3$/\textit{h}-BN}
& $P_{\mathrm{FE}}\rightarrow$ & $2.02\times10^{-5}$ & $4.19\times10^{-5}$ & $6.21\times10^{-5}$ & $1.56\times10^{-5}$ & $4.13\times10^{-4}$ & $4.29\times10^{-4}$ & 591\% \\
& $P_{\mathrm{FE}}\leftarrow$ & $3.92\times10^{-5}$ & $1.91\times10^{-5}$ & $5.83\times10^{-5}$ & $5.06\times10^{-5}$ & $3.43\times10^{-4}$ & $3.94\times10^{-4}$ & 576\% \\
\cline{2-9}
& TER & \multicolumn{3}{c}{6.5\%} & \multicolumn{3}{c}{8.9\%} & \\
\hline
\multirow{3}{*}{Gr/In$_2$Se$_3$/Gr}
& $P_{\mathrm{FE}}\rightarrow$ & $1.01\times10^{-3}$ & $1.43\times10^{-3}$ & $2.44\times10^{-3}$ & $3.35\times10^{-4}$ & $2.93\times10^{-3}$ & $3.27\times10^{-3}$ & 34\% \\
& $P_{\mathrm{FE}}\leftarrow$ & $1.29\times10^{-3}$ & $1.13\times10^{-3}$ & $2.42\times10^{-3}$ & $3.21\times10^{-4}$ & $3.07\times10^{-3}$ & $3.39\times10^{-3}$ & 40.1\% \\
\cline{2-9}
& TER & \multicolumn{3}{c}{0.8\%} & \multicolumn{3}{c}{3.7\%} & \\
\hline
\multirow{3}{*}{\textit{h}-BN/In$_2$Se$_3$/Gr}
& $P_{\mathrm{FE}}\rightarrow$ & $3.16\times10^{-4}$ & $1.79\times10^{-3}$ & $2.11\times10^{-3}$ & $5.02\times10^{-4}$ & $2.03\times10^{-3}$ & $2.53\times10^{-3}$ & 20\% \\
& $P_{\mathrm{FE}}\leftarrow$ & $2.77\times10^{-5}$ & $6.38\times10^{-5}$ & $9.15\times10^{-5}$ & $2.62\times10^{-5}$ & $8.36\times10^{-5}$ & $1.10\times10^{-4}$ & 20\% \\
\cline{2-9}
& TER & \multicolumn{3}{c}{2206\%} & \multicolumn{3}{c}{2200\%} & \\
\hline
\multirow{3}{*}{$2\cdot$\textit{h}-BN/In$_2$Se$_3$}
& $P_{\mathrm{FE}}\rightarrow$ & $2.76\times10^{-5}$ & $2.21\times10^{-5}$ & $4.97\times10^{-5}$ & $1.74\times10^{-5}$ & $8.41\times10^{-4}$ & $8.58\times10^{-4}$ & 1626\% \\
& $P_{\mathrm{FE}}\leftarrow$ & $1.05\times10^{-4}$ & $1.18\times10^{-4}$ & $2.23\times10^{-4}$ & $1.15\times10^{-4}$ & $6.91\times10^{-4}$ & $8.06\times10^{-4}$ & 261\% \\
\cline{2-9}
& TER & \multicolumn{3}{c}{349\%} & \multicolumn{3}{c}{6.5\%} & \\
\hline
\multirow{3}{*}{$2\cdot$Gr/In$_2$Se$_3$}
& $P_{\mathrm{FE}}\rightarrow$ & $1.02\times10^{-3}$ & $2.24\times10^{-3}$ & $3.26\times10^{-3}$ & $1.26\times10^{-3}$ & $8.58\times10^{-4}$ & $2.12\times10^{-3}$ & 53.8\% \\
& $P_{\mathrm{FE}}\leftarrow$ & $1.07\times10^{-3}$ & $2.13\times10^{-3}$ & $3.20\times10^{-3}$ & $1.33\times10^{-3}$ & $1.07\times10^{-3}$ & $2.40\times10^{-3}$ & 33.3\% \\
\cline{2-9}
& TER & \multicolumn{3}{c}{1.87\%} & \multicolumn{3}{c}{13.2\%} & \\
\hline\hline
\end{tabular}
}
\caption{Spin-dependent electron transmission coefficients $T_\uparrow$ and $T_\downarrow$, TMR, and TER of CrSb-based MFTJs with a Cr-symmetric (Cr-S) interface and seven different barrier layers.}
\label{table2}
\end{table*}

\renewcommand{\arraystretch}{1.3}

Before discussing the calculated transport properties, we first summarize the physical switching mechanism of the proposed altermagnetic MFTJs in Fig.~\ref{Fig2}. As shown in Fig.~\ref{Fig2}(a), the device consists of two CrSb altermagnetic electrodes, vacuum or atomically thin \textit{h}-BN/graphene insertion layers, and an $\alpha$-In$_2$Se$_3$ ferroelectric barrier. The compensated altermagnetic order of CrSb gives zero net magnetization ($M=0$), while still providing spin-split electronic states for spin-selective tunneling. The $\alpha$-In$_2$Se$_3$ barrier supplies two reversible polarization states, $P_{\rm FE}\rightarrow$ and $P_{\rm FE}\leftarrow$, which reshape the electrostatic barrier profile and band alignment. Meanwhile, the Cr/Sb termination and insertion layers provide atomic-scale interface engineering knobs for tuning the spin-dependent interfacial potential and tunneling channels.

The resulting switching mechanism is illustrated in Fig.~\ref{Fig2}(b). The device contains two coupled switching degrees of freedom: the relative N\'eel-vector configuration of the two CrSb electrodes and the ferroelectric polarization direction of the $\alpha$-In$_2$Se$_3$ barrier. Switching between the P and AP N\'eel-vector configurations changes the spin-channel matching between the two electrode/barrier interfaces and gives rise to TMR, whereas reversing the ferroelectric polarization modifies the barrier profile and produces TER. More importantly, the high- and low-resistance states are not determined by the nominal P/AP label alone, but by the actual alignment of the interfacial Cr moments. For symmetric terminations, the P configuration corresponds to antiparallel interfacial Cr moments and therefore gives a high-resistance state, whereas the AP configuration produces parallel interfacial Cr moments and a low-resistance state. For asymmetric terminations, this correspondence is reversed, so that P becomes the low-resistance configuration and AP becomes the high-resistance configuration. This termination-induced inversion persists for both $P_{\rm FE}\leftarrow$ and $P_{\rm FE}\rightarrow$, while the polarization reversal changes the barrier asymmetry and hence the TER response. Therefore, Cr/Sb termination and \textit{h}-BN or graphene insertion layers couple interfacial spin-channel matching with ferroelectric barrier reconstruction, enabling interface-controlled multistate resistance switching in CrSb/$\alpha$-In$_2$Se$_3$ altermagnetic MFTJs.

The calculated spin-resolved transmission coefficients listed in Table~\ref{table1} show that, although ferroelectric polarization reversal can modulate the interfacial transmission to some extent, both the tunneling magnetoresistance and tunneling electroresistance remain relatively small in these junctions. Unlike layered van der Waals electrodes, CrSb is a non-van-der-Waals metallic crystal, whose surface dangling bonds may strongly hybridize with the electronic states of the ferroelectric barrier. Such direct contact can modify the interfacial wave-function overlap, introduce interface states, and perturb the local electrostatic environment of $\alpha$-In$_2$Se$_3$, thereby weakening the spin-filtering effect and reducing the sensitivity of the tunneling conductance to ferroelectric polarization reversal. These results indicate that the direct CrSb/$\alpha$-In$_2$Se$_3$ interface is insufficient for achieving efficient multiferroic tunneling modulation, thus motivating the introduction of atomically thin insertion layers to regulate the interfacial coupling.

\begin{figure*}[htb!]
\centering
\includegraphics[width=17.5cm]{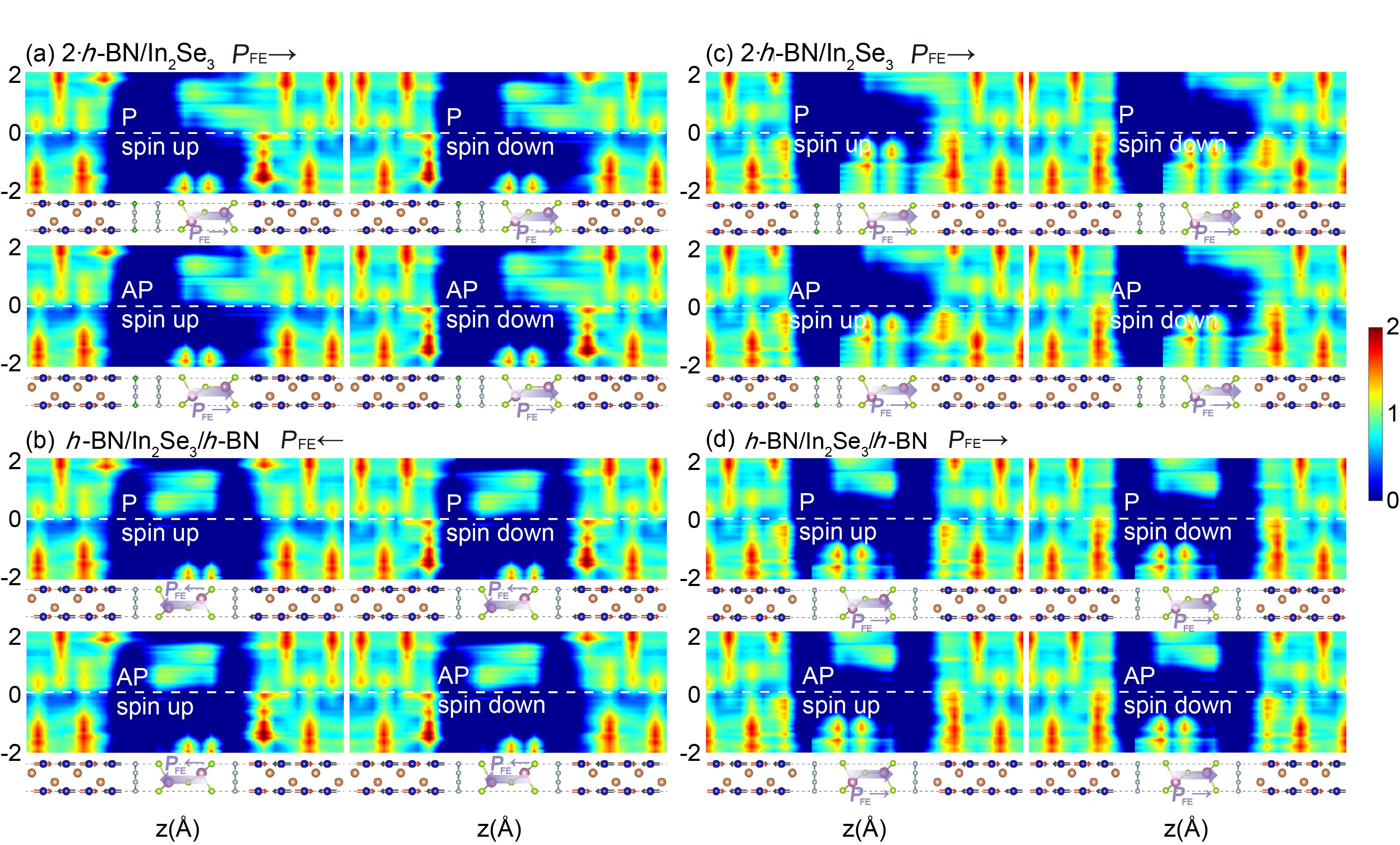}
\captionsetup{width=\textwidth}
\caption{Spin-resolved projected density of states (PDOS) along the transport direction ($z$ axis) of CrSb-based altermagnetic MFTJs with different interfaces under equilibrium conditions, together with the corresponding crystal structures of the central scattering regions. (a) P$\rightarrow$ state of the Cr-S interface; (b) P$\leftarrow$ state of the Cr-A interface; (c) P$\rightarrow$ state of the Sb-S interface; (d) P$\rightarrow$ state of the Sb-A interface. The Fermi level is indicated by the white dashed line.}
\label{Fig3}
\end{figure*}

\begin{figure*}[htb!]
\centering
\includegraphics[width=17.5cm]{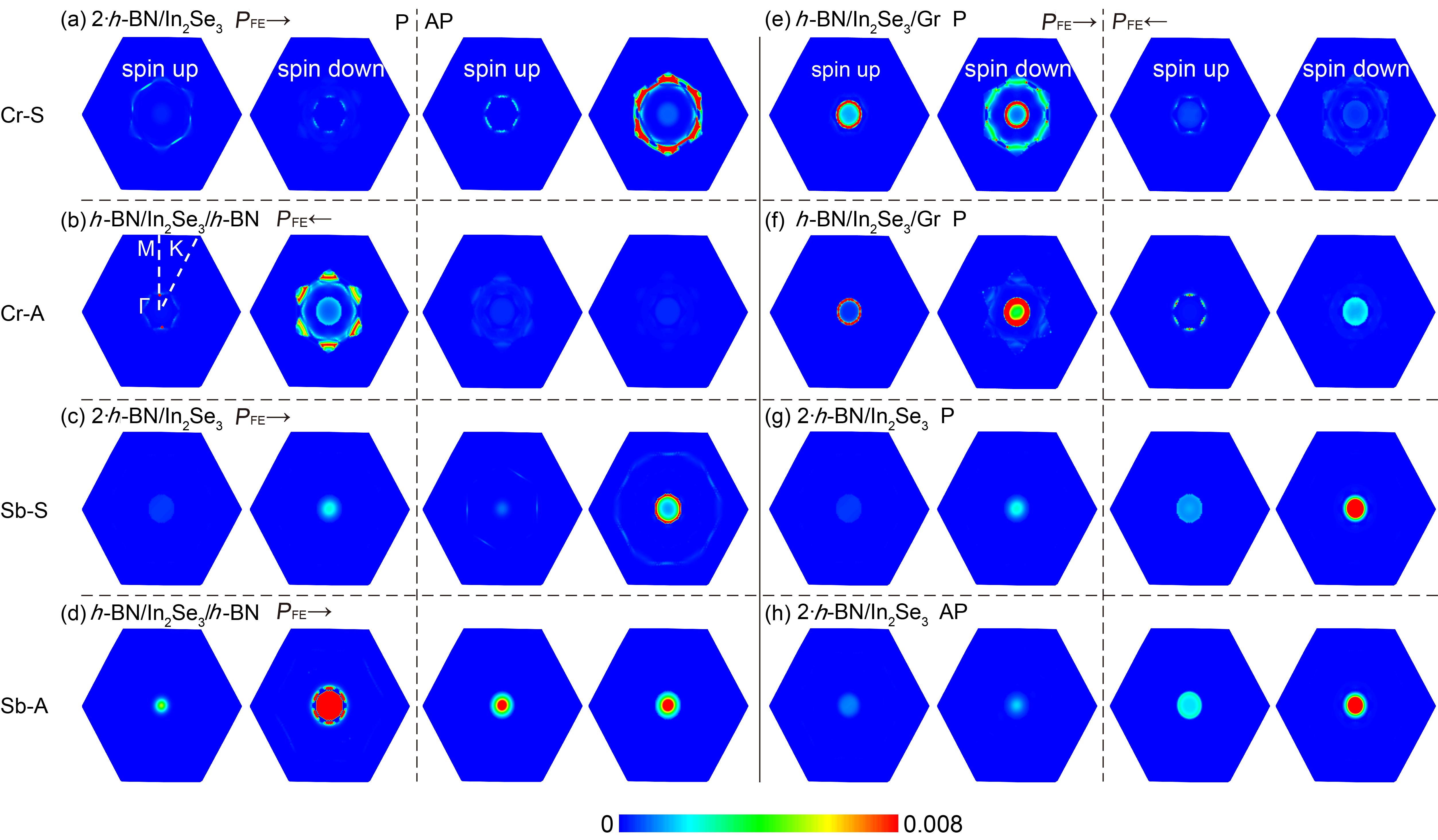}
\captionsetup{width=\textwidth}
\caption{$k_\parallel$-resolved transmission coefficients in the two-dimensional Brillouin zone at the Fermi level for CrSb-based MFTJs with four interface terminations under different ferroelectric polarization states of In$_2$Se$_3$ and P/AP N\'eel-vector configurations of CrSb. (a)(e) Cr-S interface; (b)(f) Cr-A interface; (c)(g) Sb-S interface; (d)(h) Sb-A interface.}
\label{Fig4}
\end{figure*}

To regulate this direct interfacial coupling, we introduce \textit{h}-BN and graphene (Gr) as atomically thin insertion layers between CrSb and $\alpha$-In$_2$Se$_3$. Both materials are structurally compatible with the hexagonal lattice. The optimized in-plane lattice constants are $2.50\ \mathrm{\AA}$ for \textit{h}-BN \cite{liu2003physrevb,yan2026giant}, $2.46\ \mathrm{\AA}$ for Gr \cite{dong2023physrevb}, $4.106\ \mathrm{\AA}$ for monolayer $\alpha$-In$_2$Se$_3$ \cite{jacobsgedrim2013acsnano}, and $4.101\ \mathrm{\AA}$ for CrSb. Therefore, the in-plane lattice constant of CrSb is adopted for all junction models, with $\sqrt{3}\times\sqrt{3}$ supercells of \textit{h}-BN and Gr matched to a $1\times1$ cell of $\alpha$-In$_2$Se$_3$. In this design, \textit{h}-BN serves as an electronically inert spacer to suppress excessive interfacial hybridization, whereas Gr provides a more conductive insertion layer with distinct screening and orbital-coupling characteristics. These two insertion layers therefore enable the interfacial coupling to be tuned from a weakly hybridized insulating-spacer regime to a more electronically coupled regime, providing an effective interface-engineering route to enhance the multiferroic tunneling response.

Based on this design strategy, we construct seven representative barrier configurations composed of \textit{h}-BN, Gr, and $\alpha$-In$_2$Se$_3$: (i) \textit{h}-BN/In$_2$Se$_3$, (ii) Gr/In$_2$Se$_3$, (iii) \textit{h}-BN/In$_2$Se$_3$/\textit{h}-BN, (iv) Gr/In$_2$Se$_3$/Gr, (v) \textit{h}-BN/In$_2$Se$_3$/Gr, (vi) $2\cdot$\textit{h}-BN/In$_2$Se$_3$, and (vii) $2\cdot$Gr/In$_2$Se$_3$. Following the preceding analysis of interface configurations in the direct-contact CrSb/$\alpha$-In$_2$Se$_3$/CrSb junctions, we select four representative terminations, namely Cr-S, Cr-A, Sb-S, and Sb-A, for the subsequent insertion-layer devices, in order to investigate the effects of terminal atoms and interfacial symmetry on multiferroic tunneling transport. We first optimize the interfacial stacking of Gr and \textit{h}-BN on both Cr- and Sb-terminated CrSb surfaces. As shown in Fig.~S2(a,b), the energetically preferred configurations are the Cr/Sb-hollow stacking for Gr and the Cr/Sb-N stacking for \textit{h}-BN. The optimized structures are then used to construct the two-probe MFTJ devices shown in Fig.~\ref{Fig1}(h).

\section{Interface-Controlled TMR and TER at Equilibrium}

We first evaluate the equilibrium TMR and TER of monolayer-In$_2$Se$_3$-based CrSb MFTJs with four CrSb interface terminations and seven barrier configurations. As summarized in Table~\ref{table2} for the representative Cr-S interface and in Tables~S1--S3 for the other terminations, the transport response depends sensitively on both the Cr/Sb termination and the barrier composition. For monolayer-In$_2$Se$_3$ barriers, Cr-terminated interfaces (Cr-S and Cr-A) generally show stronger resistance modulation than Sb-terminated interfaces (Sb-S and Sb-A), indicating that interfacial Cr atoms provide more favorable spin-channel matching for altermagnetic tunneling. This trend demonstrates that the resistance modulation is not determined by the ferroelectric barrier alone, but by the combined effect of interface termination and insertion-layer composition.

To analyze the microscopic origin of the resistance contrast, we select representative junctions that give the largest TMR or TER for each interface termination. For the Cr-S interface, the maximum TMR reaches 1626\% in the $2\cdot$\textit{h}-BN/In$_2$Se$_3$ barrier, while the maximum TER reaches 2206\% in the asymmetric \textit{h}-BN/In$_2$Se$_3$/Gr barrier. For the Cr-A interface, the optimal TMR and TER are 575\% and 1992\%, respectively. In contrast, Sb-terminated interfaces show weaker modulation, with maximum TMR and TER values below 500\%. These representative cases are used below to clarify how the local interfacial magnetic arrangement, spin-channel matching, and polarization-dependent barrier profile determine the high- and low-resistance states.

\begin{figure*}[htb!]
\centering
\includegraphics[width=\textwidth]{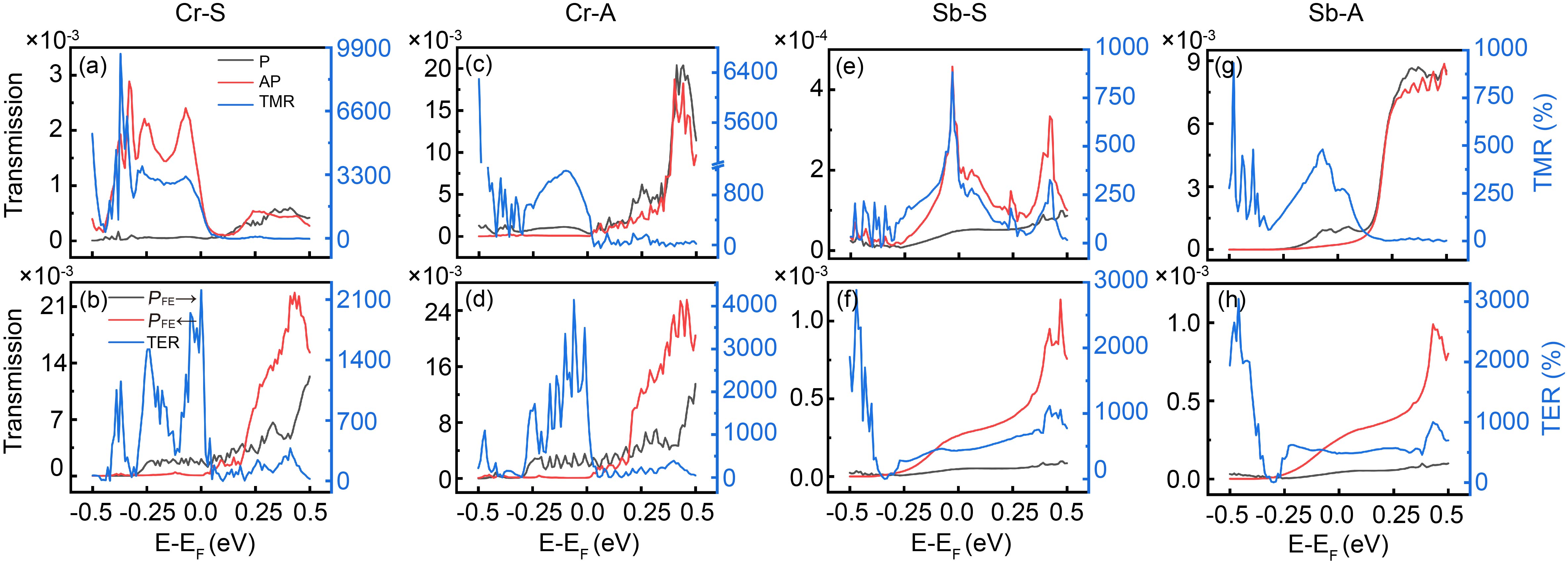}
\captionsetup{width=\textwidth}
\caption{Transmission coefficient near the Fermi level of the MFTJ with P and AP configurations, and the corresponding TMR and TER ratios as a function of energy relative to the Fermi level ($E-E_\text{F}$). The left vertical axis represents the transmittance, and the right vertical axis represents TMR (\%) and TER (\%).}
\label{Fig5}
\end{figure*}

\begin{figure*}[htb!]
\centering
\includegraphics[width=17.5cm]{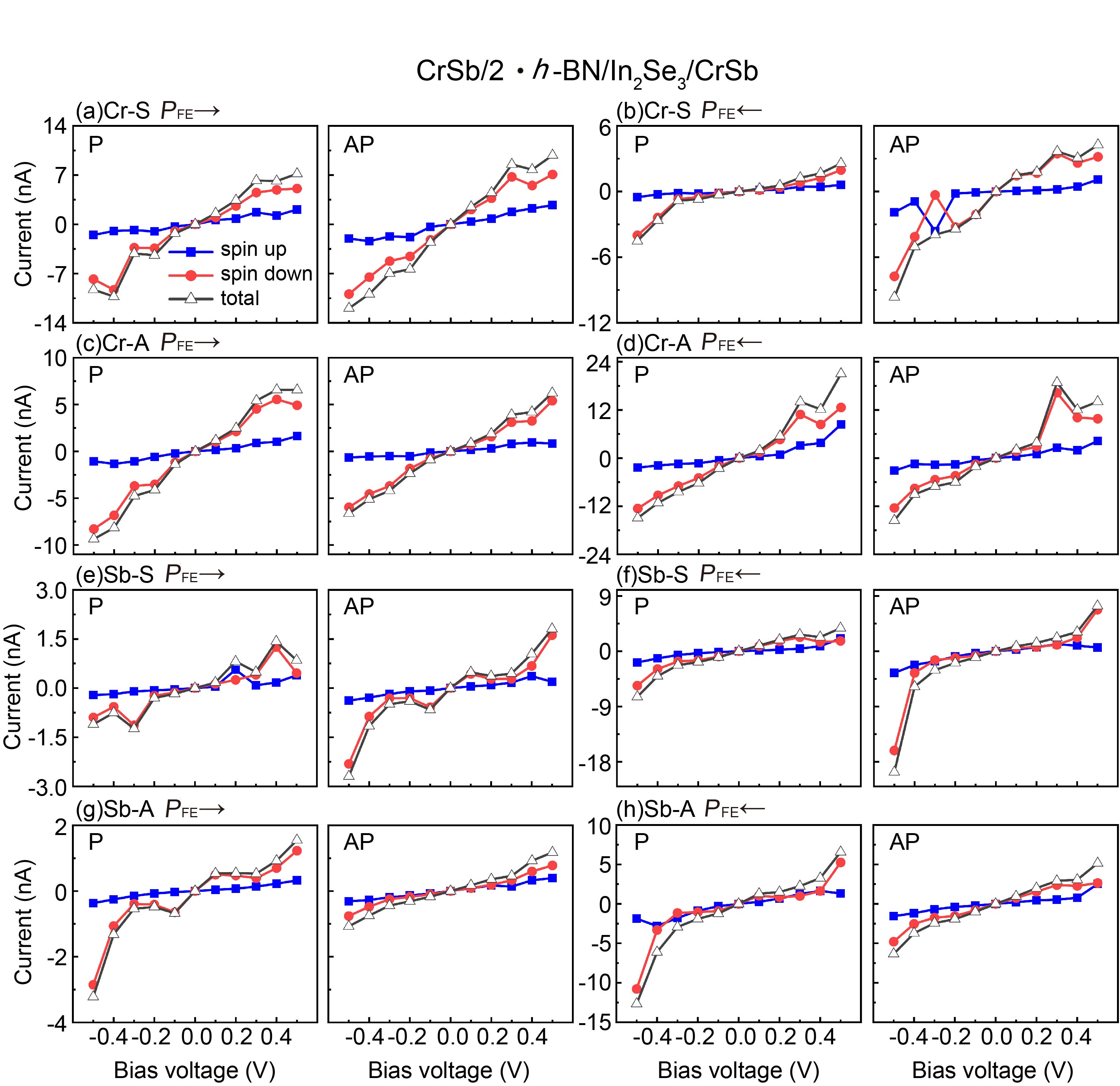}
\captionsetup{width=\textwidth}
\caption{Current as a function of bias voltage for the CrSb/$2\cdot$\textit{h}-BN/In$_2$Se$_3$/CrSb multiferroic tunnel junction with four different interfaces; (a-b) Cr-S interface; (c-d) Cr-A interface; (e-f) Sb-S interface; (g-h) Sb-A interface.}
\label{Fig6}
\end{figure*}

The microscopic origin of the high- and low-resistance states can be understood from the spin-resolved real-space projected density of states (PDOS) in the central scattering region, plotted in the $(E,z)$ plane in Fig.~\ref{Fig3}. For all four interface types, the spectral weight at the Fermi level is strongly suppressed inside the barrier region, confirming that transport occurs in the tunneling regime. More importantly, the interfacial magnetic arrangement determines which spin channel maintains better spectral continuity across the junction. For the symmetric interfaces (Cr-S and Sb-S), the P alignment of the electrode N\'eel vectors corresponds to an antiferromagnetic arrangement of the interfacial Cr moments, which suppresses tunneling at the Fermi level and produces a high-resistance state. When the right-electrode N\'eel vector is switched to the AP configuration, the interfacial Cr moments become ferromagnetically aligned, leading to stronger spectral continuity across the barrier for the dominant spin channel and hence a low-resistance state. For the asymmetric interfaces (Cr-A and Sb-A), the correspondence between the P/AP alignment and the interfacial Cr-moment arrangement is reversed, and the high- and low-conductance states are accordingly interchanged, as shown in Fig.~\ref{Fig3}(b,d). Therefore, the TMR is not determined simply by the nominal P or AP label, but by the actual magnetic alignment of the interfacial Cr moments and its matching with the spin-dependent tunneling channels.

This picture is further supported by the $k_{\parallel}$-resolved transmission at the Fermi level in the two-dimensional Brillouin zone, as shown in Fig.~\ref{Fig4}. Figures~\ref{Fig4}(a)--(d) correspond to the junctions with the maximum TMR for the four interface terminations. For the symmetric interfaces (Cr-S and Sb-S), transmission hot spots are concentrated mainly in the spin-down channel of the low-resistance configuration, whereas they are strongly suppressed in the corresponding high-resistance state. This behavior indicates not only a large TMR but also a pronounced spin-filtering effect. The asymmetric interfaces exhibit the reversed trend, consistent with the reversed relation between P/AP alignment and interfacial Cr-moment arrangement. In addition, Figs.~\ref{Fig4}(e)--(h) show that reversing the ferroelectric polarization of $\alpha$-In$_2$Se$_3$ significantly redistributes the transmission weight in momentum space. The clear change in the hot-spot distribution between P$\rightarrow$ and P$\leftarrow$ demonstrates that the ferroelectric barrier does not merely shift the overall resistance, but actively reshapes the available tunneling channels. The momentum-resolved transmission therefore provides direct evidence for the coexistence of large TMR and TER in the proposed MFTJs under equilibrium conditions.

Finally, we examine the energy dependence of the optimized CrSb-based MFTJs within the range from $-0.5$ to $+0.5$ eV relative to the Fermi level. As shown in Fig.~\ref{Fig5}, both the transmission and the resulting TMR/TER ratios exhibit strong energy dependence. For the Cr-S interface, the TMR reaches 9576\% at $-0.37$ eV [Fig.~\ref{Fig5}(a)], while for the Cr-A interface the TER increases to 4144\% at $-0.06$ eV [Fig.~\ref{Fig5}(e)]. These off-Fermi-level enhancements indicate that the multiferroic transport response is highly sensitive to the energy-dependent interfacial tunneling spectrum. Therefore, the giant TMR and TER are not restricted to a single equilibrium energy point, but can be further modulated when the effective transport window is shifted, for example by electrostatic gating, carrier doping, or finite-bias operation. This energy dependence suggests an additional route for controlling the spin- and polarization-dependent tunneling characteristics of CrSb-based MFTJs.

\begin{figure*}[htb!]
\centering
\includegraphics[width=17.5cm]{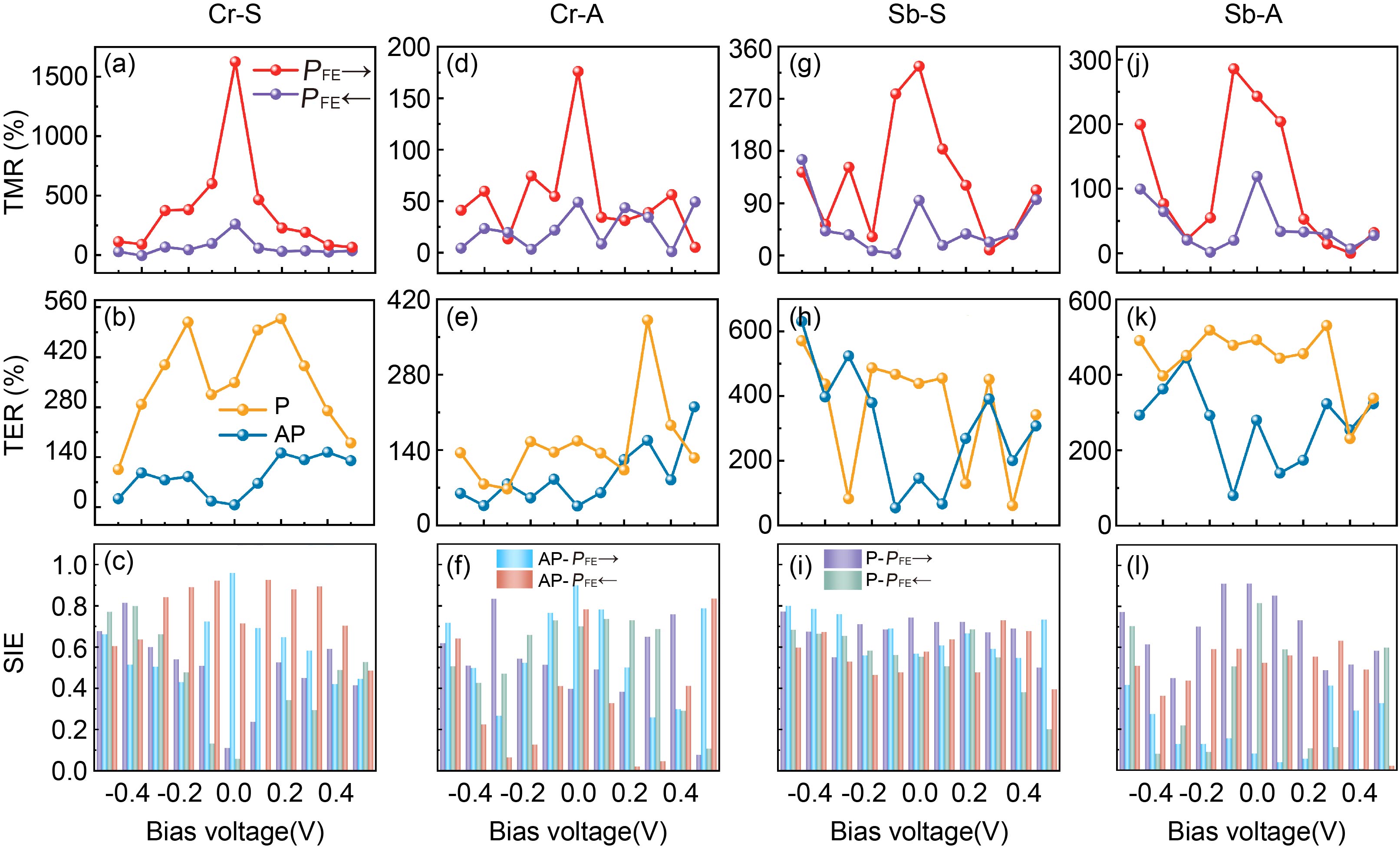}
\captionsetup{width=\textwidth}
\caption{TMR (panels a, d, g, j), TER (panels b, e, h, k), and SIE (panels c, f, i, l) as a function of bias voltage for the CrSb/$2\cdot$\textit{h}-BN/In$_2$Se$_3$/CrSb multiferroic tunnel junction with four different interfaces; (a-c) Cr-S interface; (d-f) Cr-A interface; (g-i) Sb-S interface; (j-l) Sb-A interface.}
\label{Fig7}
\end{figure*}

\subsection{Spin-Polarized Transport under Nonequilibrium Conditions}

To examine whether the interface-controlled multiferroic response survives beyond equilibrium, we calculate the bias-dependent spin-resolved current, TMR, TER, and spin-injection efficiency (SIE) in the range from $-0.5$ to $0.5$ V. The main text focuses on the CrSb/$2\cdot$\textit{h}-BN/In$_2$Se$_3$/CrSb junction, which exhibits the largest equilibrium TMR among the monolayer-In$_2$Se$_3$-based structures. Results for the CrSb/\textit{h}-BN/In$_2$Se$_3$/\textit{h}-BN/CrSb and CrSb/\textit{h}-BN/In$_2$Se$_3$/Gr/CrSb junctions are provided in Figs.~S3--S6 of the Supporting Information.

Figure~\ref{Fig6} shows the calculated $I$--$V$ curves of the CrSb/$2\cdot$\textit{h}-BN/In$_2$Se$_3$/CrSb junction with four interface terminations. Overall, the magnitude of the total current increases with increasing bias for most magnetic and ferroelectric states, consistent with the widening of the bias-induced transport window. Nevertheless, the current remains strongly dependent on interface termination, N\'eel-vector alignment, and ferroelectric polarization. The spin-resolved currents show clear spin asymmetry in most configurations, indicating that spin-selective tunneling persists under finite bias. This spin asymmetry is especially pronounced for the Cr-S interface, where one spin channel dominates the total current over a broad bias range. The $I$--$V$ curves also exhibit bias-polarity asymmetry, particularly for polarization-reversed states, reflecting the combined effect of the polar ferroelectric barrier and the asymmetric interfacial electrostatic potential.

\begin{figure*}[htb!]
\centering
\includegraphics[width=17cm]{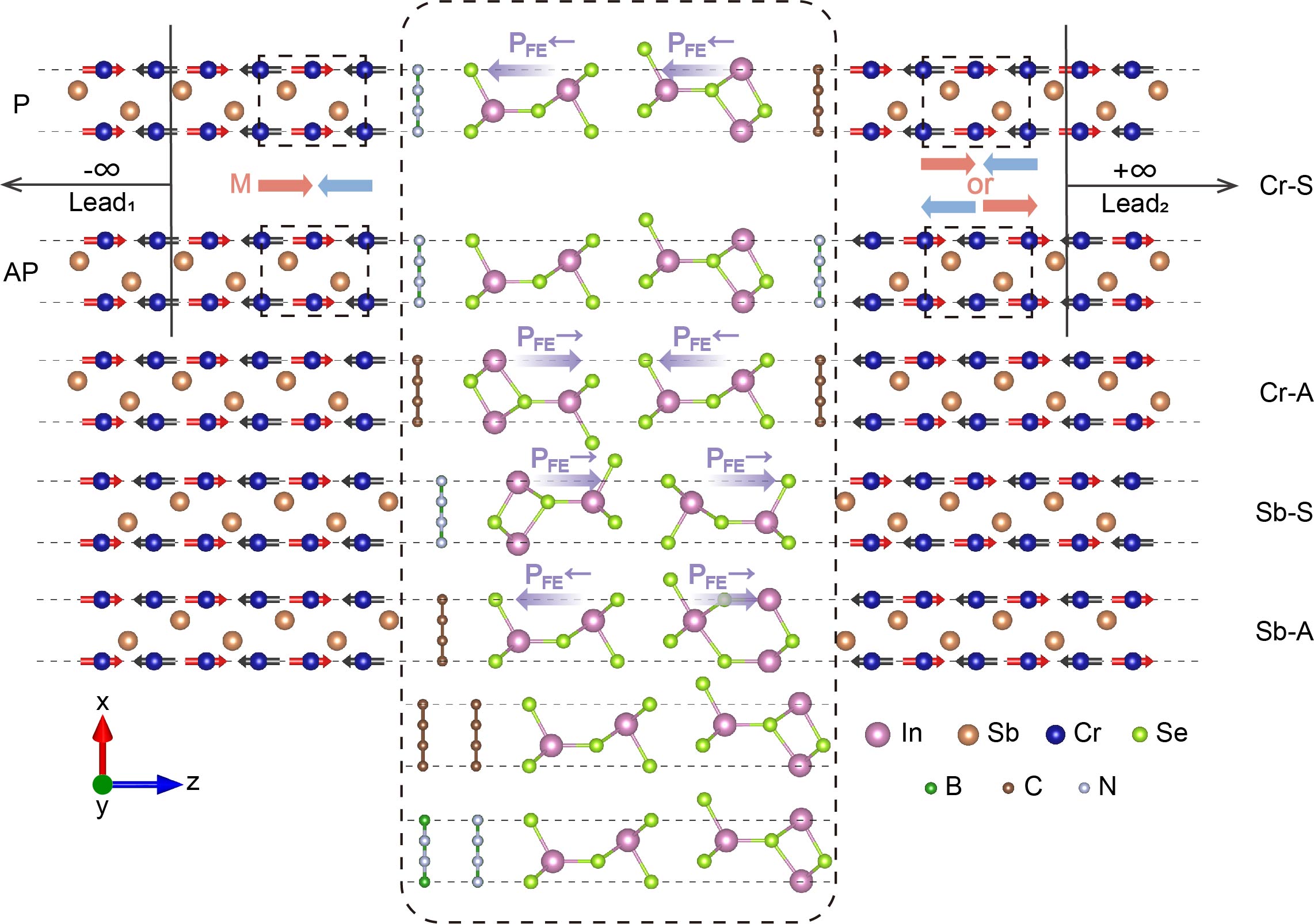}
\captionsetup{width=\textwidth}
\caption{Structural schematics of multiferroic tunnel junctions with bilayer In$_2$Se$_3$ as the ferroelectric barrier, showing four different interface termination types and seven distinct barrier layer configurations. The periodic unit cells of the left and right electrodes are highlighted with black dashed rectangles. The arrows on the Cr atoms indicate the local magnetic moments. The $z$-axis is the transport direction, and periodicity is maintained within the $xy$-plane.}
\label{Fig8}
\end{figure*}

The corresponding TMR, TER, and SIE ratios are summarized in Fig.~\ref{Fig7}. Unlike the total current, the TMR and TER show pronounced nonmonotonic bias dependence, indicating that the resistance ratios are controlled by the overlap between the bias window and the spin- or polarization-dependent transmission channels. For the Cr-S interface, a large TMR is retained near equilibrium, reaching about 1626\% for the P$\rightarrow$ polarization state [Fig.~\ref{Fig7}(a)]. The TER response shows a stronger dependence on polarization-induced barrier asymmetry and can remain several hundred percent over a broad bias range for selected terminations, particularly Sb-S and Sb-A [Figs.~\ref{Fig7}(h,k)]. This behavior indicates that the optimal interface for TMR and TER need not be the same under finite bias.

The additional finite-bias results in the Supporting Information show that the barrier composition further modulates the balance between current amplitude, spin filtering, and ferroelectric resistance switching. The CrSb/\textit{h}-BN/In$_2$Se$_3$/\textit{h}-BN/CrSb junction exhibits sharp TMR peaks at selected bias voltages (Figs.~S3 and S4), while the graphene-inserted CrSb/\textit{h}-BN/In$_2$Se$_3$/Gr/CrSb junction generally supports larger current amplitudes and enhanced TER peaks (Figs.~S5 and S6). These trends are consistent with the reduced effective tunneling resistance and stronger interfacial electronic coupling introduced by the thinner spacer configuration or by graphene insertion.

In addition to the resistance ratios, the finite-bias SIE provides useful information on spin selectivity. For the CrSb/$2\cdot$\textit{h}-BN/In$_2$Se$_3$/CrSb junction, the SIE remains sizable over a broad bias range for all four interface terminations [Figs.~\ref{Fig7}(c,f,i,l)]. This is consistent with the spin-resolved $I$--$V$ curves in Fig.~\ref{Fig6}, where one spin channel often dominates the total current. Therefore, the spin-selective tunneling induced by the altermagnetic CrSb electrodes is not limited to equilibrium, but can survive under finite-bias operation.

These nonequilibrium results demonstrate that CrSb-based altermagnetic MFTJs retain sizable TMR, TER, and spin-polarized current responses under finite bias. The bias voltage does not simply suppress the resistance ratios; instead, it reshapes the effective transport window and can selectively enhance TMR or TER when favorable spin- or polarization-dependent transmission channels are sampled. The distinct behaviors of different barrier configurations further show that interface insertion layers provide an effective means to balance current amplitude, spin filtering, and ferroelectric resistance modulation in altermagnetic multiferroic tunnel junctions.

\subsection{Equilibrium Transport Properties Modulated by Bilayer In$_2$Se$_3$ Barriers}

To further extend the multistate functionality of CrSb-based multiferroic tunnel junctions and explore the role of interlayer ferroelectric polarization coupling in spin-dependent tunneling, we replace the monolayer In$_2$Se$_3$ barrier with bilayer In$_2$Se$_3$. As shown in Fig.~\ref{Fig8}, bilayer In$_2$Se$_3$ can host four stable polarization configurations, namely the up, down, head-to-head, and tail-to-tail states. When these four ferroelectric configurations are combined with the parallel (P) and antiparallel (AP) alignments of the CrSb N\'eel vectors, the junction can in principle generate eight nonvolatile resistance states. This provides an additional degree of freedom for multistate memory and electric/magnetic cooperative modulation.

We first calculated the bilayer In$_2$Se$_3$-barrier tunnel junctions without insertion layers. The calculated spin-resolved equilibrium transmission coefficients, together with the corresponding TMR and TER values, are summarized in Table~S4. Compared with the monolayer In$_2$Se$_3$ junctions, the bilayer In$_2$Se$_3$ barrier exhibits larger TMR and TER, indicating that the additional ferroelectric layer provides stronger modulation of spin-dependent tunneling transport. However, the transport performance still strongly depends on the interface termination, suggesting that the direct CrSb/In$_2$Se$_3$ contact continues to play an important role in spin-dependent tunneling. Therefore, it is necessary to further introduce \textit{h}-BN and Gr insertion layers to regulate the interfacial coupling.

Based on this design concept, we construct bilayer-In$_2$Se$_3$-based CrSb MFTJs with four interface terminations, namely Cr-S, Cr-A, Sb-S, and Sb-A, consistent with the monolayer-barrier devices. In addition, \textit{h}-BN and Gr are introduced as insertion layers to form seven representative composite barriers, denoted as \textit{h}-BN/Gr, $2\cdot$\textit{h}-BN, \textit{h}-BN/\textit{h}-BN, \textit{h}-BN, $2\cdot$Gr, Gr/Gr, and Gr, respectively, as illustrated in Fig.~\ref{Fig8}. This model series enables a systematic comparison of how interface termination, barrier composition, and bilayer ferroelectric polarization configuration affect the equilibrium TMR and TER under the same CrSb altermagnetic electrodes.

\begin{figure*}[htb!]
\centering
\includegraphics[width=17.5cm]{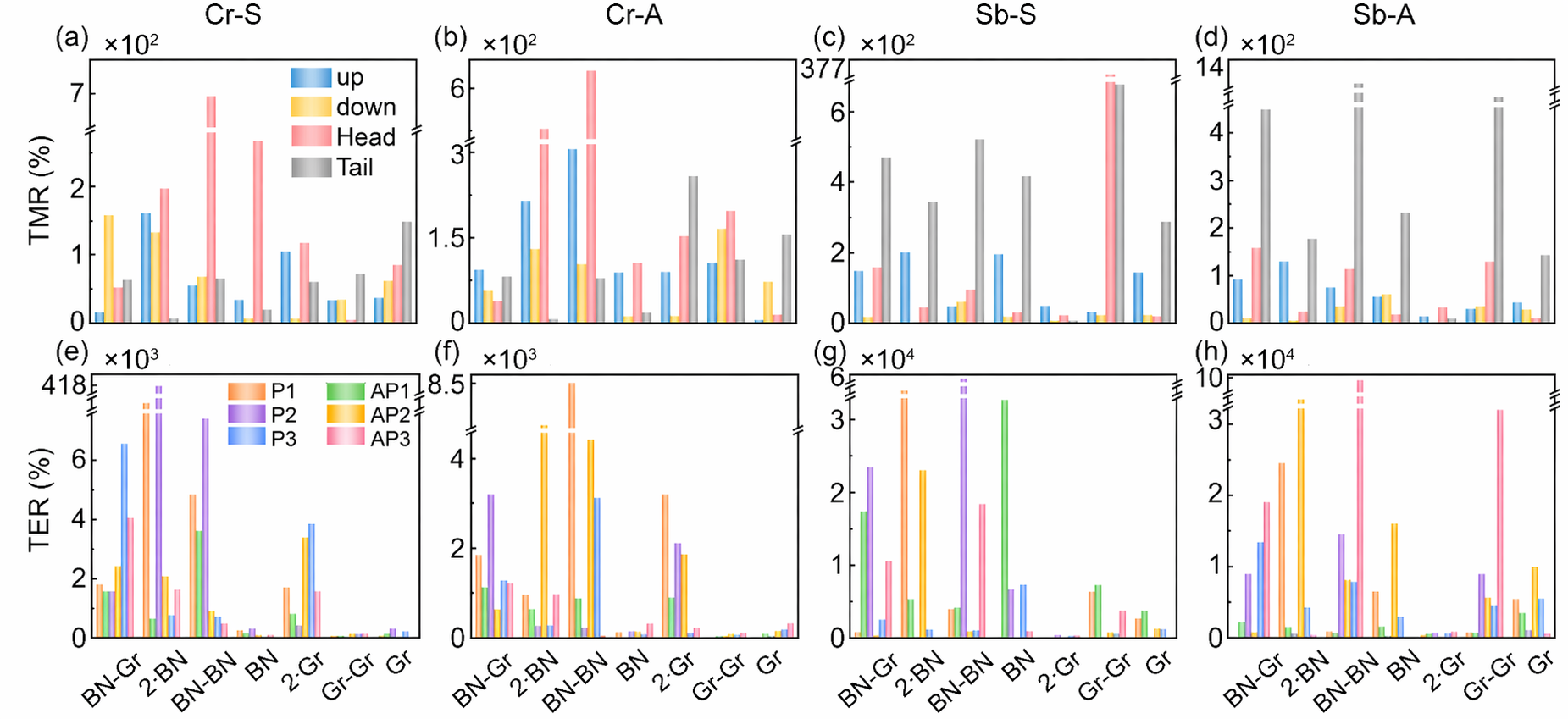}
\captionsetup{width=\textwidth}
\caption{TMR and TER of the CrSb multiferroic tunnel junction with bilayer In$_2$Se$_3$ under four different interface terminations, for seven different barrier layers. (a)(e) Cr-S interface; (b)(f) Cr-A interface; (c)(g) Sb-S interface; (d)(h) Sb-A interface.}
\label{Fig9}
\end{figure*}

Figures~\ref{Fig9}(a)--(d) show the equilibrium TMR of the bilayer-In$_2$Se$_3$-based CrSb MFTJs with four different interface terminations. The TMR exhibits a strong dependence on interface termination, barrier composition, and ferroelectric polarization configuration. For most interface types, changing the bilayer polarization state leads to a sizable variation in TMR, indicating that the interlayer polarization arrangement can effectively modify the spin-dependent tunneling channels. Among all configurations considered, the CrSb/Gr/bilayer-In$_2$Se$_3$/Gr/CrSb junction with the Sb-S interface shows the largest TMR, reaching $3.77 \times 10^{4}\%$ [Fig.~\ref{Fig9}(c)]. This result differs from the monolayer-barrier case, where Cr-terminated interfaces generally dominate, and indicates that bilayer polarization configurations introduce an additional electrostatic degree of freedom that can overcome the simple termination trend. In this case, the combined effect of graphene insertion, Sb-S termination, and bilayer polarization reshapes the band matching and spin-dependent transmission across the barrier, strongly amplifying the conductance difference between the P and AP states.

The TER enhancement induced by bilayer In$_2$Se$_3$ is even more pronounced. Figures~\ref{Fig9}(e)--(h) present the TER values obtained by comparing different pairs of ferroelectric polarization states under the P and AP magnetic configurations for the four interface terminations. The TER generally reaches the order of $10^{3}\%$ in many configurations, and several cases are further enhanced to the order of $10^{4}\%$ or even $10^{5}\%$. In particular, for the CrSb/$2\cdot$\textit{h}-BN/bilayer-In$_2$Se$_3$/CrSb junction with the Cr-S interface, the TER reaches a maximum value of $4.18 \times 10^{5}\%$ [Fig.~\ref{Fig9}(e)]. This ultrahigh TER indicates that different bilayer polarization configurations can lead to drastically different tunneling conductances, providing a highly effective route for resistance-state modulation.

The large TER can be understood from the strong sensitivity of tunneling transport to the polarization-dependent potential landscape. Compared with monolayer In$_2$Se$_3$, bilayer In$_2$Se$_3$ provides not only two uniform polarization states, but also head-to-head and tail-to-tail configurations with distinct internal electrostatic profiles. These polarization configurations can modify the effective barrier height, interfacial charge screening, and band alignment between the electrodes and the barrier. Consequently, the transmission probability can vary strongly among different polarization states, giving rise to much larger TER values than in the monolayer case. These results show that bilayer ferroelectric barriers provide an effective strategy for increasing both the number of nonvolatile resistance states and the magnitude of resistance modulation in CrSb-based altermagnetic MFTJs.

\section{CONCLUSIONS}

In summary, we have investigated CrSb-based altermagnetic MFTJs with different Cr/Sb terminations, insertion layers, and ferroelectric barrier configurations, and clarified the microscopic origin of their spin- and polarization-dependent tunneling. The large resistance modulation is not simply a consequence of the bulk altermagnetic spin splitting of CrSb or the ferroelectric polarization reversal of In$_2$Se$_3$, but arises from the coupling among symmetry selected interface termination, spin channel matching, and ferroelectric barrier reshaping. For monolayer-In$_2$Se$_3$ barriers, Cr-terminated interfaces generally yield stronger resistance modulation than Sb-terminated ones, with representative TMR and TER values reaching 1626\% and 2206\%, respectively; moreover, symmetric and asymmetric terminations induce an inversion of the P/AP resistance hierarchy by reconfiguring the dominant spin-resolved tunneling channels. The insertion of \textit{h}-BN or graphene further modifies the interfacial potential, orbital hybridization, and spin-channel matching, while energy-dependent transmission and finite-bias calculations show that the resistance contrast, spin filtering, and spin-polarized current can be tuned by shifting the effective transport window. Extending the barrier to bilayer In$_2$Se$_3$ introduces interlayer polarization coupling, increases the number of nonvolatile resistance states from four to eight, and gives maximum TMR and TER values of $3.77\times10^{4}\%$ and $4.18\times10^{5}\%$, respectively. These results identify interface symmetry, termination-dependent spin-channel matching, and ferroelectric barrier reconstruction as coupled design principles for stray-field-free, electrically switchable, multistate spintronic tunnel devices.

\section{COMPUTATIONAL METHODS}

All structural relaxations, total-energy calculations, and electronic-structure calculations were performed within density functional theory (DFT) using the Vienna \textit{Ab initio} Simulation Package (VASP) \cite{kresse1996physrevb}. The projector augmented-wave (PAW) method was used to describe the electron--ion interaction \cite{bloechl1994physrevb}, and the exchange--correlation functional was treated within the generalized gradient approximation (GGA) using the Perdew--Burke--Ernzerhof (PBE) form \cite{perdew1996physrevlett}. Long-range van der Waals interactions were included using the DFT-D3 method. A plane-wave cutoff energy of 500 eV was adopted. For slab and periodic heterostructure models, a vacuum region of 20 \AA{} was introduced along the out-of-plane direction to avoid spurious interactions between periodic images. All atoms were fully relaxed until the total-energy convergence criterion reached $10^{-6}$ eV and the residual forces were smaller than 0.01 eV/\AA{}. A $\Gamma$-centered $13\times13\times1$ $k$-point mesh was used for structural optimization and total-energy calculations of the two-dimensional heterostructures, while a $12\times12\times8$ mesh was used for bulk CrSb. We did not apply an on-site Hubbard correction to CrSb because the electronic structure calculated with $U=0$ shows the best agreement with available angle-resolved photoemission spectroscopy (ARPES) measurements \cite{yang2025natcommun,ding2024physrevlett}. Unless otherwise specified, spin-polarized calculations were performed within a collinear magnetic framework without spin--orbit coupling because the altermagnetic spin splitting in CrSb is primarily exchange-driven.

Electron transport properties were calculated by combining DFT with nonequilibrium Green's function (NEGF) theory, as implemented in the Nanodcal package \cite{taylor2001physrevb1,taylor2001physrevb2}. In the self-consistent transport calculations, a double-$\zeta$ polarized basis set was used for the atomic orbitals \cite{soler2002jphyscond}, the kinetic-energy cutoff was set to 80 Hartree, and the Hamiltonian convergence threshold was set to $10^{-4}$ eV. The electronic temperature was fixed at 300 K through the Fermi--Dirac distribution. To ensure convergence of the transmission spectra and current, a $100\times100\times1$ $k$-point mesh was employed for all CrSb-based MFTJ configurations.

Within the NEGF--DFT formalism, the spin-polarized current $I_\sigma$ and conductance $G_\sigma$ are given by the Landauer--B\"uttiker expressions \cite{meir1992physrevlett,datta1995mesoscopic}
\begin{equation}
I_\sigma = \frac{e}{h} \int T_\sigma(E)\left[f_L(E)-f_R(E)\right]\mathrm{d}E,
\end{equation}
\begin{equation}
G_\sigma = \frac{e^2}{h}T_\sigma(E_F),
\end{equation}
where $\sigma=\uparrow,\downarrow$ denotes the spin channel, $e$ is the electron charge, $h$ is Planck's constant, $T_\sigma(E)$ is the spin-resolved transmission coefficient, and $f_L(E)$ and $f_R(E)$ are the Fermi--Dirac distribution functions of the left and right electrodes, respectively. The total current is
\begin{equation}
I = I_\uparrow + I_\downarrow.
\end{equation}
The spin-injection efficiency (SIE) is defined as
\begin{equation}
\mathrm{SIE} = \left|\frac{I_\uparrow-I_\downarrow}{I_\uparrow+I_\downarrow}\right|.
\end{equation}
At zero bias, the corresponding quantity can be evaluated from the transmission coefficients at the Fermi level.

The tunneling magnetoresistance (TMR) at equilibrium is defined as \cite{yuasa2004natmater}
\begin{equation}
\begin{aligned}
\mathrm{TMR}
&=\frac{|G_{\mathrm{P}}-G_{\mathrm{AP}}|}
{\min\{G_{\mathrm{P}},G_{\mathrm{AP}}\}}\times100\% \\
&=\frac{|T_{\mathrm{P}}-T_{\mathrm{AP}}|}
{\min\{T_{\mathrm{P}},T_{\mathrm{AP}}\}}\times100\% ,
\end{aligned}
\end{equation}
where $G_{\mathrm{P}}$ and $G_{\mathrm{AP}}$ (or $T_{\mathrm{P}}$ and $T_{\mathrm{AP}}$) denote the conductances (or total transmission coefficients at the Fermi level) for the parallel and antiparallel N\'eel-vector configurations of the two CrSb electrodes, respectively. Under a finite bias voltage $V$, the TMR is evaluated from the magnitudes of the currents as
\begin{equation}
\mathrm{TMR}(V)=
\frac{\left|\,|I_{\mathrm{P}}(V)|-|I_{\mathrm{AP}}(V)|\,\right|}
{\min\{|I_{\mathrm{P}}(V)|,|I_{\mathrm{AP}}(V)|\}}\times100\% .
\end{equation}

The tunneling electroresistance (TER) is defined by comparing the two opposite ferroelectric polarization states of the barrier \cite{tao2016applphyslett,kang2021physrevb},
\begin{equation}
\begin{aligned}
\mathrm{TER}
&=\frac{|G_{\rightarrow}-G_{\leftarrow}|}
{\min\{G_{\rightarrow},G_{\leftarrow}\}}\times100\% \\
&=\frac{|T_{\rightarrow}-T_{\leftarrow}|}
{\min\{T_{\rightarrow},T_{\leftarrow}\}}\times100\% ,
\end{aligned}
\end{equation}
where $G_{\rightarrow}$ and $G_{\leftarrow}$ (or $T_{\rightarrow}$ and $T_{\leftarrow}$) denote the conductances (or total transmission coefficients at the Fermi level) for the two opposite ferroelectric polarization states of the barrier under a fixed magnetic configuration. At a finite bias voltage $V$, the TER under a fixed magnetic configuration $m$ ($m=\mathrm{P}$ or $\mathrm{AP}$) is defined as

\begin{equation}
\mathrm{TER}_{m}(V)
=
\frac{
\left|\,|I_{m,\rightarrow}(V)|-|I_{m,\leftarrow}(V)|\,\right|
}{
\min\{|I_{m,\rightarrow}(V)|,|I_{m,\leftarrow}(V)|\}
}
\times 100\%.
\end{equation}
Here, $I_{m,\rightarrow}$ and $I_{m,\leftarrow}$ are the total currents for the two opposite polarization states under the same magnetic configuration $m$ and applied bias.

\begin{acknowledgement}
This work was supported by the National Natural Science Foundation of China Regional Innovation and Development Joint Fund Key Program (No.~U24A6002) and the National Natural Science Foundation of China (No.~12304148).
\end{acknowledgement}

\bibliography{sample}

\end{document}